\documentclass[twocolumn,amsmath,amssymb]{snp}
\usepackage{epsfig}
\usepackage{hyperref}
\usepackage{url}
\usepackage{graphicx}
\usepackage{dcolumn}
\usepackage{bm}
\usepackage{graphicx,amssymb}
\usepackage{color}
\usepackage{float}

\usepackage{multirow}
\usepackage{rotating}
\usepackage{lineno}
\usepackage{subfigure}
\begin{document}

\title{\Large Charmonium production in pp collisions at energies available at the CERN Large Hadron Collider}

\author{Biswarup Paul}
\email{biswarup.paul@saha.ac.in}
\author{ Mahatsab Mandal}
\author{ Pradip Roy}
\author{Sukalyan Chattapadhyay}
\affiliation{Saha Institute of Nuclear Physics, 1/AF Bidhannagar
Kolkata - 700064, India}
\maketitle
The quarkonia production mechanism in $p-p$ collisions is qualitatively understood by the models based on the Quantum Chromodynamics(QCD), in particular, in the non-relativistic QCD (NRQCD)~\cite{prd51}. The quarkonia production in NRQCD is a two step process: First, the creation of the $Q\bar Q$ pair in a hard scattering at short distances which are process-dependent, is to be calculated perturbatively as an expansions in the strong coupling constant $\alpha_{s}$. Note that $Q\bar Q$ states can be in the color-singlet(CS)~\cite{zpc19} as well as in a color-octet(CO)~\cite{prl74} states. Second, the $Q\bar Q$ pair evolves into the quarkonium state with the probabilities that are given by the supposedly universal nonperturbative long-distance matrix elements (LDMEs) which have to be extracted from experiments. For CO states, this evolution process also involves the nonperturbative emission of soft gluons to form a CS states. The crucial feature of this formalism is that the complete structure of the $Q\bar Q$ Fock space, which is spanned by the states $n=\,^{2S+1}L^{[i]}_J$ with spin $S$, orbital angular momentum $L$, total angular momentum $J$, and color multiplicity $i=1,8$.
The inclusive cross-section of charmoium include the prompt contribution (the sum of direct production and feeddown contributions from the decay of heavier charmonium states) and the B feeddown. The prompt contribution has been calculated using above mentioned NRQCD formalism and 
FONLL~\cite{jhep9805} formalism is dedicated to calculate the feeddown contributions from B meson to the $J/\psi$ and $\psi(2S)$ productions.

According to the NRQCD
factorization  formalism, the cross section for direct production 
of a resonance $H$ in a collision of particle $A$ and $B$ can be expressed as 
\begin{eqnarray}
&&d\sigma_{A+B\rightarrow H+X} = \sum_{i,j,n}\int dx_a dx_b  G_{a/A}(x_a,\,\mu_F)\nonumber\\
&&\, G_{b/B}(x_b,\,\mu_F) d\sigma(a+b\rightarrow Q\bar Q(n) +X)<\mathcal{O}^H(n)>
\end{eqnarray}
where $G_{a/A}(G_{b/B})$ is the partonic distribution function(PDF) of the 
incoming partons $a(b)$ in the incident hadron $A(B)$ which depends on 
the  large light-cone momentum fraction $x_a(x_b)$ and the factorization 
scale $\mu_F$. The short distance contribution $d\sigma(a+b\rightarrow Q\bar Q(n) +X)$ 
can be calculated within the framework of 
perturbative QCD(pQCD). On the other hand, $<\mathcal{O}^H(n)>$ are nonperturbative 
LDMEs and to be extracted from experiment. 
\begin{figure}
\includegraphics[scale=0.30]{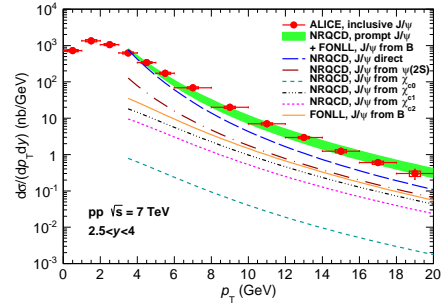}
\vspace{-3.25mm}
\caption{\label{fig1} Differential production cross-section vs. $p_{T}$ for inclusive $J/\psi$ compared with the ALICE data~\cite{epjc}.}
\end{figure}
\begin{figure}
\includegraphics[scale=0.30]{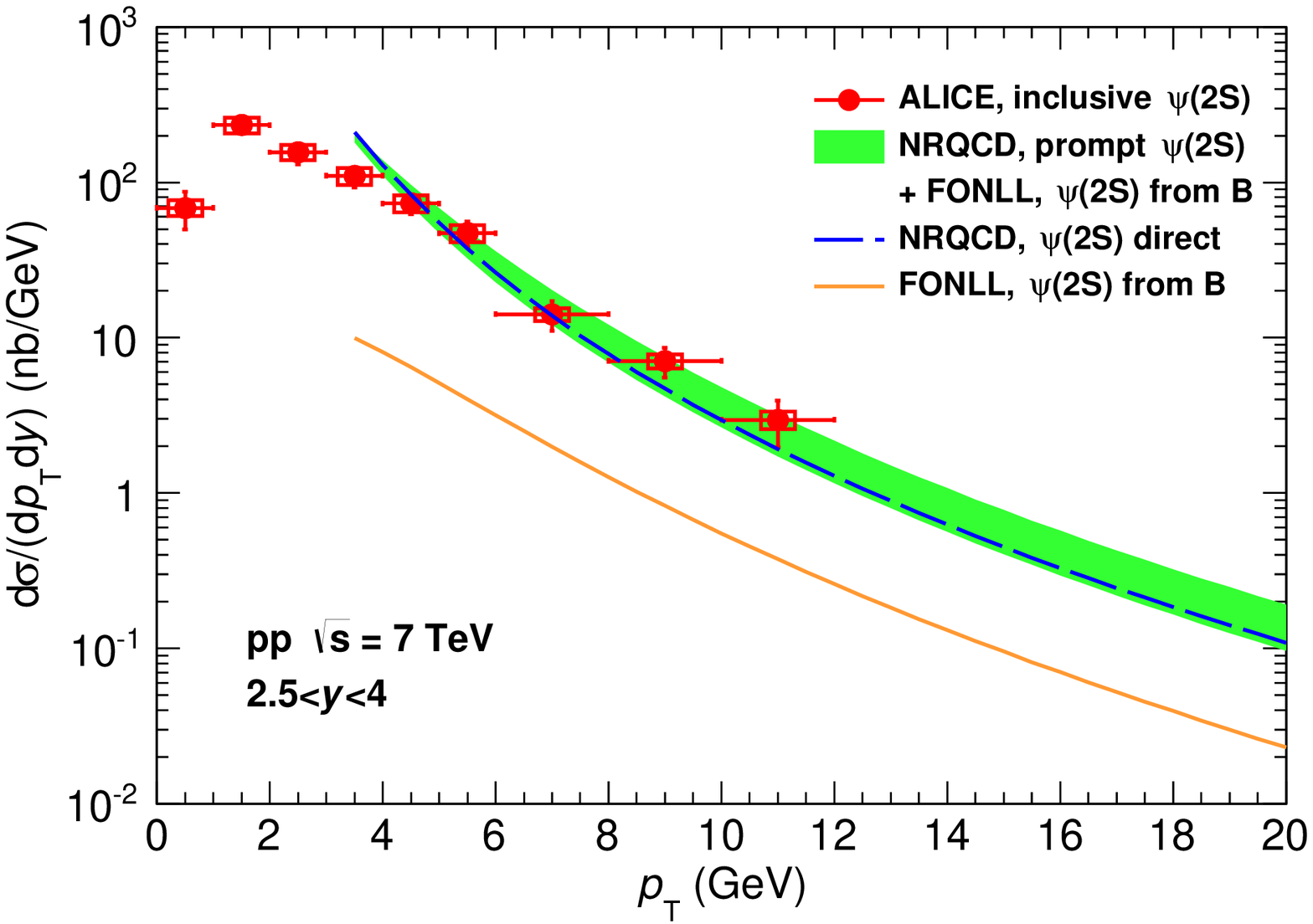}
\vspace{-3.25mm}
\caption{\label{fig2} Same as Fig. \ref{fig1} but for inclusive $\psi(2S)$.}
\end{figure}
\begin{figure}
\includegraphics[scale=0.30]{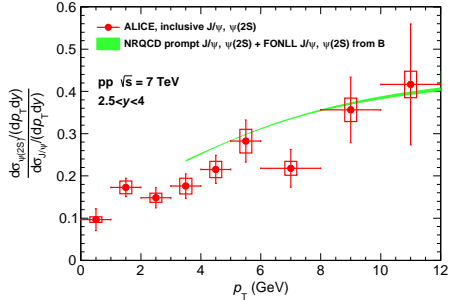}
\vspace{-3.25mm}
\caption{\label{fig3} Inclusive $\psi(2S)$ to $J/\psi$ production cross-section ratio vs. $p_{T}$ compared to the ALICE data~\cite{epjc}.}
\end{figure}
\begin{figure}
\includegraphics[scale=0.30]{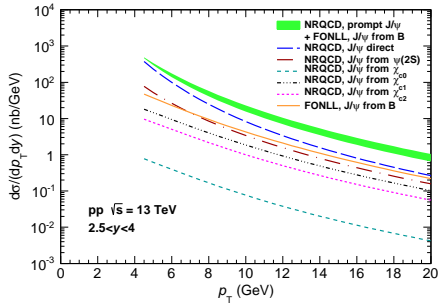}
\vspace{-3.25mm}
\caption{\label{fig4} Same as Fig. \ref{fig1} but only with the theoretical prediction for the differential cross-sections for inclusive $J/\psi$ at $\sqrt{s}$ = 13 TeV~\cite{prdR}.}
\end{figure}
\begin{figure}
\includegraphics[scale=0.30]{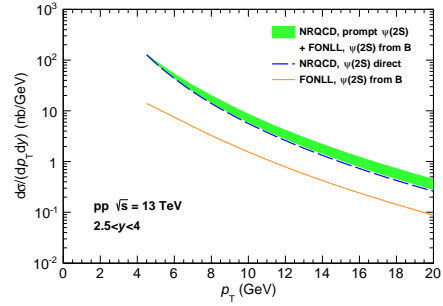}
\vspace{-3.25mm}
\caption{\label{fig5} Same as Fig. \ref{fig4} but for inclusive $\psi(2S)$.}
\end{figure}  
\section*{Results}
The shaded band in Fig.~\ref{fig1} and ~\ref{fig2} represent the predictions for the differential cross-sections of inclusive $J/\psi$ and $\psi(2S)$ as a function of $p_{T}$ for the rapidity interval 2.5$<$y$<$4 at $\sqrt{s}$ = 7 TeV, respectively, within the NRQCD and FONLL framework~\cite{prdR}. 
The uncertainty limits on the calculated values correspond to the variation of the factorization scale $\mu_{F}$. This uncertainty due to the factorization scale was estimated by performing the calculations for $\mu_{F}$ = $\mu_{R}$ =$\mu_{0}$/2 (upper bound) and $\mu_{F}$ = $\mu_{R}$ = 2$\mu_{0}$ (lower bound), here $\mu_{0}$ = $\sqrt{p_T^2 + 4m_c^2}$. These uncertainties were calculated for the direct production as well as for the feeddown contributions from $\psi(2S)$, $\chi_{c0}$, $\chi_{c1}$ and $\chi_{c2}$. However, in the figure, the feeddown contributions have been shown by lines which correspond to the calculated values for the central values of LDMEs and $\mu_{F}$. The experimental data from ALICE~\cite{epjc} for the inclusive $J/\psi$ and $\psi(2S)$ differential cross-sections has also been shown in this figures. The NRQCD calculations are unable to describe the experimental data for $p_{T} <$ 4 GeV since in this domain the perturbative approximation fails. ALICE collaboration has also reported the $\psi(2S)$ to $J/\psi$ cross-sections ratio at $\sqrt{s}$ = 7 TeV~\cite{epjc}. These measured and calculated values are shown in Fig.~\ref{fig3}. The experimental data shows a clear dip for the bin 6-8 GeV while the calculated values show a smooth increasing trend in the domain of 3.5 $< p_{T} <$ 12 GeV. The predicted cross-section for $J/\psi$ and $\psi(2S)$ for $\sqrt{s}$ = 13 TeV are shown in Figs.~\ref{fig4} and \ref{fig5}, respectively.

\end{document}